\documentclass[superscriptaddress,groupedaddress,12pt]{article}  
\usepackage{graphicx}  
\usepackage{dcolumn}   
\usepackage{bm}        
\usepackage{amssymb}   
\usepackage{jcappub}

\usepackage{bm}
\usepackage{latexsym}
\usepackage{amsmath}
\usepackage{amsfonts}
\usepackage{multirow}
\usepackage{array}
\usepackage{rotating}
\usepackage{epstopdf}
\usepackage{subfigure}

\begin{document}

\title{Screening fifth forces in k-essence and DBI models}

\author[1]{Philippe Brax,}
\author[2]{Clare Burrage}
\author[3]{and Anne-Christine Davis}
\affiliation[1]{Institut de Physique Th\'{e}orique, CEA, IPhT, CNRS, URA2306, F-91191
Gif-sur-Yvette c\'{e}dex, France}
\affiliation[2]{School of Physics and Astronomy, University of Nottingham, Nottingham NG7 2RD, UK}
\affiliation[3]{Department of Applied Mathematics and Theoretical Physics,
Centre for Mathematical Sciences, Cambridge CB3 0WA, UK}

\emailAdd{Philippe.Brax@cea.fr}
\emailAdd{Clare.Burrage@nottingham.ac.uk}
\emailAdd{A.C.Davis@damtp.cam.ac.uk}

\abstract{New fifth forces have not yet been detected in the laboratory or in the solar system, hence it is typically difficult to introduce new light scalar fields that would mediate such forces.  In recent years it has been shown that a number of non-linear scalar field theories allow for a dynamical mechanism, such as the Vainshtein and  chameleon ones, that suppresses the strength of the scalar fifth force in experimental environments.  This is known as screening, however it is unclear how common screening is within non-linear scalar field theories.    k-essence models are commonly studied  examples of non-linear models, with DBI as the best motivated example, and so  we ask whether these non-linearities are able to screen a scalar fifth force. We  find that a Vainshtein-like screening mechanism exists for such models although with limited applicability. For instance, we cannot find a screening mechanism for DBI models. On the other hand, we construct a large class of k-essence models which lead to the acceleration of the Universe in the recent past for which the fifth force mediated by the scalar can be screened.
 }

\maketitle

\section{Introduction}

The study of non-linear scalar field theories has rapidly expanded in recent years, because the non-linearities enable the scalar field to be compatible with observations when related linear fields theories would be excluded\footnote{   In this paper we call ``linear" any model where  scalar fields have equations of motion in which terms depend on only one power of the scalar field, all other theories will be called ``non-linear".}.  Particular success has been had in constructing theories which are compatible with the null results of searches for fifth forces.  Experimental searches for fifth forces have been conducted in the laboratory, and the leading results come from the E\"{o}t-Wash group \cite{Adelberger:2009zz}. The presence of fifth forces can also be constrained by studying the motion of bodies in the solar system \cite{2002nmgm.meet.1797W}.  Neither of these extremely precise searches have detected any evidence for additional forces, yet if a new scalar field is introduced into a theory we expect it to mediate a new force.

The simplest kind of scalar field we could imagine introducing is a canonical scalar with a mass term. In our terminology this is a linear theory.   If such a field is light and  couples to matter then the fifth force experiments we mentioned above constrain the strength of this interaction to be at least five orders of magnitude weaker than gravity. This implies that the energy scale controlling the strength of the interaction must be five orders of magnitude above the Planck scale.  This is clearly very difficult to justify in the context of a sensible effective field theory.

There are a number of ways we could attempt to avoid this problem.  We could insist that new scalar fields must be sufficiently heavy that they cannot mediate forces over the distance scales probed in these experiments.  However attempts to explain the late time acceleration of the expansion of the universe typically require very light scalar degrees of freedom.  We could insist that these new light scalar fields are forbidden from interacting with matter fields, for example by imposing a shift symmetry on the scalar field that would forbid the problematic coupling.  However when studying such axionic models it is necessary to break the shift symmetry in order to reproduce the desired cosmology, and depending on how this breaking occurs the problematic coupling to matter may be reintroduced \cite{Brax:2009kd,Panda:2010uq}.  The most successful attempts to introduce new light scalar fields which couple to matter without violating the results of fifth force experiments rely on making the theory non-linear.

Non-linear terms can be introduced into the potential for the scalar field \cite{Khoury:2003rn}, the way it couples to matter \cite{Hinterbichler:2010es}, or the kinetic terms \cite{Nicolis:2008in}. If the non-linearities are irrelevant in vacuum, but become important in the presence of sufficient quantities of matter  then they change the way the field interacts with that matter distribution.  This can result in the force being suppressed in experimental environments  without the need to fine tune any parameters.  Considering  these theories from a classical point of view, the role of  the non-linearities  is to impede the growth of the scalar potential well around the massive source.  If the scalar potential well is shallower than the gravitational well the scalar force will be weaker than the gravitational force.  From a particle theory perspective it is possible to see why the scalar force is suppressed by  perturbing the scalar field around a background configuration $\phi_0$. Then to second order, the Lagrangian for the fluctuations $\delta \phi$ has the following form
\begin{equation}
\mathcal{L}\supset -\frac{Z(\phi_0)}{2}(\partial\delta\phi)^2+\frac{m^2(\phi_0)}{2}\delta\phi^2+\frac{\beta (\phi_0)}{M_P}\delta\phi \delta T\;,
\label{eq:pertlag}
\end{equation}
where the final term describes the interactions between the scalar field fluctuations and the trace of the energy momentum tensor of matter fluctuations.   If the background scalar field configuration $\phi_0$ is non-trivial due to the presence of a source then  the non-linearities present in the theory can act to make  the coefficient of the kinetic term, $Z$, the mass of the field, $m$, or the strength of its coupling to matter, $\beta$, vary between different points in space and the scalar force may be screened. If $Z(\phi_0)$  increases  the propagation of scalar fluctuations is impeded, thus making the scalar force difficult to transmit, this is the essence of the Vainshtein screening mechanism \cite{Vainshtein:1972sx,Deffayet:2001uk} that is employed by Galileon models with non-linear kinetic terms \cite{Nicolis:2008in}.  If $m(\phi_0)$ increases   the scalar force only  propagates over shorter distances, this is the essence of the chameleon screening mechanism \cite{Khoury:2003rn}.  If $\beta(\phi_0)$ decreases  the interaction between the scalar field and matter weakens, this is the essence of the symmetron screening mechanism \cite{Hinterbichler:2010es} or the Damour-Polyakov phenomenon for dilaton models \cite{Brax:2010gi,Brax:2011ja,Damour:1994zq}.

It is currently unclear whether most non-linear theories contain screening mechanisms, or whether the models that have been studied are unusual exceptions.  The non-linearities could cause the functions in the Lagrangian of Equation (\ref{eq:pertlag}) to change in the opposite way to that required, thus increasing the strength of the force, or the non-linearities could cause changes in more than one function that conspire to cancel out the effects of screening.  In this paper we study a well motivated class of non-linear theories and ask whether the presence of these non-linearities allows the scalar force to be screened.  These are known as k-essence models.    The k-essence Lagrangian  is an arbitrary function of $\phi$ and $X=(1/2)(\partial \phi)^2$.  These Lagrangians  were first introduced in the context of kinetically driven inflation \cite{ArmendarizPicon:1999rj,Garriga:1999vw}, and were later applied to dark energy \cite{Chiba:1999ka,ArmendarizPicon:2000dh,ArmendarizPicon:2000ah}.  Despite the presence of Lagrangian terms with many derivatives k-essence models are free of Ostrogradski's ghost  and can be considered as a specific case of the most general ghost free scalar field theory first identified by Horndeski \cite{Horndeski:1974,Deffayet:2009mn}.  A particular example is the DBI Lagrangian \cite{Aharony:1999ti};
\begin{equation}
\mathcal{L} =-T(\phi)\sqrt{1+\frac{(\partial\phi)^2}{T(\phi)}}-V(\phi) +T(\phi) +\frac{\phi}{M}\rho\;.
\end{equation}
This is derived from string theory and describes the open string dynamics of D-branes. The DBI Lagrangian is believed to be UV complete and is therefore taken as
more physically motivated than other k-essence theories.
A proof that Vainshtein-like screening solutions can exist for generalised Galileons, which includes k-essence models, was given in \cite{Shirai:2012iw}.  In this paper we specialise to k-essence and ask specifically what form a k-essence model must take in order to have a successful screening mechanism.

In the following Section we will review the screening mechanisms that are currently known.  Then in Section \ref{sec:kess} we will introduce k-essence models and their cosmology and  describe the approach we will take to studying screening in k-essence models.  In Section \ref{sec:chamlike} we study whether a chameleon-like screening mechanism exists for k-essence.  In Section \ref{sec:vainsteinoneterm} we study whether a Vainshtein-like screening mechanism exists for k-essence.  We give an example of a model that screens in this way in Section \ref{sec:example}, and we are able to extend our analysis to the case when the k-essence Lagrangian contains more than one operator in Section \ref{sec:multiple} and then to a fully non-perturbative treatment in Section \ref{sec:non-pert}.  In Section \ref{sec:nonpert} we show that these models can also give rise to a cosmology that matches the one we observe.  We conclude in Section \ref{sec:conclusions}.

\section{Known screening mechanisms}
We will take a classical approach to studying whether the scalar force is screened, by computing the scalar field profile around a massive source and from this deducing the scalar force.  We will assume that quantum corrections can be safely neglected.  In this section we review  the screening mechanisms that are currently known from this standpoint.  An important practical distinction between models will be whether the non-linearities become important only inside a massive source, as in the first two examples we discuss, or whether they become important exterior to the source as in the last example.

In this article we will only consider theories in which the coupling between the scalar field and matter fields takes the form
\begin{equation}
\mathcal{L} \supset A(\phi)T\;,
\end{equation}
where $T$ is the trace of the energy momentum tensor of the matter fields\footnote{  Other types of coupling are possible \cite{Bekenstein:1992pj}, and also lead to interesting behaviour, however in this article we restrict ourselves to the most commonly studied form of coupling.}.  These  scalar fields are referred to as having conformal couplings to matter and scalars coupled in this way are the most difficult to reconcile with experimental searches for fifth forces.

The force mediated by a scalar field that couples in this way is
\begin{equation}
\vec{F}=\frac{\beta}{M_P}\vec{\nabla}\phi\;,
\end{equation}
where
\begin{equation}
\beta(\phi)=M_P \frac{\partial A(\phi)}{\partial\phi}\;.
\end{equation}
For static, spherically symmetric scenarios the scalar force will only be non-trivial in a radial direction.

\subsection{A linear theory}
We begin by showing what happens in the linear case, so that it is easier to see how the non-linear models with screening mechanisms differ. The linear Lagrangian is
\begin{equation}
\mathcal{L}_{\phi}= -\frac{1}{2}(\partial \phi)^2 - \frac{1}{2}m^2\phi^2+\frac{\beta\phi}{M_P}T\;.
\end{equation}
The simplest scenario in which to study scalar forces is to consider a spherical, static massive source of density $\rho$ and radius $R$, corresponding to a mass $M_c=(4/3)\pi R^3\rho$.  Here and in what follows we assume that the background spacetime is flat and that the gravitational effects of the source can be well described with a Newtonian theory.   Then inside the source the scalar field grows as
\begin{equation}
\phi=\frac{\beta\rho}{m^2M_P}\left(1-\frac{\sinh mr}{mr}\right)\;,
\end{equation}
where, because we are interested in light scalar fields, we have made the approximation $mR\ll 1$.  Outside the source
\begin{equation}
\phi=-\frac{\beta M_c}{4\pi M_p r}e^{-mr}\;.
\end{equation}
Therefore outside the source the ratio of the scalar to Newtonian forces is
\begin{equation}
\left|\frac{F_{\phi}}{F_N}\right|=2\beta^2 (1+mr)e^{-mr}\;.
\end{equation}
From this is is clear that on distance scales shorter than the Compton wavelength of the scalar field, $mr\ll 1$, the scalar force can only be suppressed by tuning $\beta$ to be small.

\subsection{The chameleon}
The archetypal chameleon model starts from the Lagrangian
\begin{equation}
\mathcal{L}_{\phi}=-\frac{1}{2}(\partial \phi)^2 -\Lambda^4\left(\frac{\Lambda}{\phi}\right)+\frac{\beta\phi}{M_p}T\;,
\end{equation}
although many other choices of potential are possible.  The role of the potential is to make the effective mass of the scalar field much larger inside a massive non-relativistic  source of density $\rho$ than outside, this is because the scalar field dynamics are governed by an effective potential
\begin{equation}
V_{\rm eff}(\phi)=\frac{\Lambda^5}{\phi}+\frac{\beta\phi\rho}{M_P}\;.
\end{equation}
The mass of the field at the minimum of the potential is
\begin{equation}
m^2=2\Lambda^5\left(\frac{\beta\rho}{\Lambda^5 M_P}\right)^{3/2}\;.
\label{eq:chammass}
\end{equation}
Clearly the mass of the field changes with the local value of $\rho$.

How to solve the chameleon equations of motion has been described in detail in \cite{Khoury:2003rn,Mota:2006fz}, but the important features of the solution are that inside a sufficiently  massive object the scalar field takes a constant value everywhere except very close to the surface.  This is how we see that the scalar potential well is shallower for the chameleon, than for a corresponding linear scalar field,  because the growing mass of the scalar field makes it increasingly costly for the scalar field to decrease its value.  Outside the massive source, and at distances smaller than the Compton wavelength of the field $rm_{\infty}\ll 1$,  the ratio of the chameleon scalar force to the Newtonian one is
\begin{equation}
\frac{F_{\phi}}{F_N}=\frac{\beta(\phi_{\infty}-\phi_c)}{M_P}\frac{8\pi M_P^2 R}{M_c}\;,
\end{equation}
where $\phi_c=\sqrt{\Lambda^5/M_P\beta\rho}$ and $\phi_{\infty}=\sqrt{\Lambda^5/M_P\beta\rho_{\infty}}$ are the values of the scalar field that minimise the effective potential inside and outside the object respectively. From this expression it is clear that the scalar force is weaker than gravity if the depth of the scalar potential well inside the source, $\beta\Delta\phi/M_P=\beta(\phi_{\infty}-\phi_c)/M_P$, is shallower that the corresponding gravitational potential well, $\Phi=M_c/8\pi M_P^2 R$.

\subsection{The symmetron}
The original symmetron model begins with the Lagrangian
\begin{equation}
\mathcal{L}_{\phi}=-\frac{1}{2}(\partial\phi)^2+\frac{1}{4\lambda}(\mu^2-\lambda\phi^2)^2-\frac{\phi^2}{2M^2}T\;.
\end{equation}
As with the chameleon model the position of the minimum of the effective scalar potential varies depending on the local energy density.  When this energy density is sufficiently high $\rho>M^2\mu^2$ the minimum is at $\phi=0$.  At this minimum the coupling between scalar fluctuations and matter vanishes, and so the scalar field no-longer responds to the presence of a massive source.  In the interior of such dense objects the scalar potential well stops growing, and so the scalar force sourced by that object is suppressed.  Outside the same source object of density $\rho$ and radius $R$ that we considered before, and inside the Compton wavelength of the field $\mu r \ll 1$,  the ratio of the symmetron scalar force to the Newtonian one is
\begin{equation}
\frac{F_{\phi}}{F_N}=\frac{\mu^2}{M^2\lambda}\frac{8\pi M_P^2 R}{M_c}\left(1-\frac{R}{r}\right)\;.
\end{equation}
At infinity the minimum of the effective potential for the scalar field is at $\phi_{\infty}=\mu/\sqrt{\lambda}$ and within a sufficiently massive object the minimum is at $\phi_c=0$, so as in the chameleon case we see that the scalar force is suppressed compared to gravity whenever the potential and form of the coupling to matter conspire to stop the scalar potential well, $\Delta \phi^2/M^2$, from growing within the source object as fast as the gravitational well grows, $\Phi=M_c/8\pi M_P^2 R$.

\subsection{The Galileon and the Vainshtein mechanism}
\label{sec:vainshtein}
The final screening mechanism that is known behaves somewhat differently from the two previously discussed.  The simplest Lagrangian that displays this form of screening is
\begin{equation}
\mathcal{L}_{\phi}=-\frac{1}{2}(\partial\phi)^2-\frac{1}{2\Lambda^3}(\partial\phi)^2\Box\phi +\frac{\beta\phi}{M_P}T\;.
\end{equation}
In this example the non-linearities are gradient terms, and therefore they can become important in the exterior of the source object.  The Galileon theory shows how such higher order derivative interactions can be constructed without giving rise to ghosts.  The non-linearities dominate the behaviour of the scalar field within a  specific radius, known as the Vainshtein radius;
\begin{equation}
R_V=\left(\frac{\beta M_c}{2\pi M_P}\right)^{1/3}\frac{1}{\Lambda}\;.
\end{equation}
On solving the equations of motion we find that within this radius the ratio of the scalar to Newtonian forces is
\begin{equation}
\frac{F_{\phi}}{F_N}=2\beta^2\left(\frac{r}{R_V}\right)^{3/2}\;.
\end{equation}
Outside the Vainshtein radius the scalar force is unscreened, but within it the force is suppressed, and the further inside the radius the more the force is suppressed.

In the discussion of screening mechanisms for k-essence models that follows we will consider two possibilities, one that the non-linearities may only be important inside the source, we will call these chameleon-like screening mechanisms, and second that the non-linearities can be important outside the source object, we will call these Vainshtein-like screening mechanisms.

\section{k-essence models}
\label{sec:kess}
We will consider an operator expansion of the k-essence Lagrangian, $\mathcal{L}(\phi, X)$.  It may be that to describe a particular theory completely requires an infinite series of operators, however in any given situation we expect only a finite set of operators to dominate the behaviour of the field, and we are interested in which of these operators allow for screening.  We assume that the scalar field couples to matter conformally. Such a coupling is always present unless there is a symmetry to forbid it, and this  is not the case for a general k-essence scenario.   Therefore our starting point is the scalar Lagrangian
\begin{equation}
\mathcal{L}_{\phi}=\sum_{n,m}M_{m,n\geq 0}^{4-n-4m}\phi^nX^m -\frac{\beta}{M_P}\phi T\;,
\label{eq:startlag}
\end{equation}
where
\begin{equation}
X=\frac{1}{2}(\partial\phi)^2\;,
\end{equation}
and $T$ is the trace of the energy momentum tensor for matter.  We assume that the coupling to matter and the non-linearities of the theory are controlled by different mass scales; $M_P/\beta$ and $M_{n,m}$ respectively.

The simplest scenario in which to study the existence of absence of screening is a spherically symmetric, static non-relativistic source of density $\rho$, radius $R$ and mass $M_c$.  Therefore $T=-\rho(r)=-\rho \Theta(R-r)$.  Working around a flat space background  the resulting equation of motion is
\begin{equation}
\sum_{n,m}M_{m,n}^{4-n-4m}n \phi^{n-1}X^m +\frac{\beta}{M_P}\rho(r)=\frac{1}{\sqrt{-g}}\partial_{\mu}\left[\sum_{m,n}\sqrt{-g}M_{m,n}^{4-n-4m}m\phi^n\partial^{\mu}\phi X^{m-1}\right]\;,
\end{equation}
where $\sqrt{-g}$ is not automatically trivial, for instance when spherical coordinates are used.

\subsection{k-essence screening}

Our approach in what follows will be based on the examples of known screening mechanisms discussed in the previous section.  We will assume that there is  a near vacuum region in which the non-linearities of the theory are unimportant.  If a massive source object is introduced into this vacuum then we expect the non-linear terms of the theory to be excited in the neighborhood of the source, and  provide a dynamical screening mechanism.  As mentioned before we will consider two cases separately one where the non-linear terms are excited outside the source object, which we will call Vainshtein-like, and then other where they are only excited inside, which we will call chameleon-like.

We will solve the equations of motion in a piecewise fashion,  breaking  space  into regions inside and outside the object and regions where the non-linearities are important and un-important.  We will impose that the scalar field and its derivative are continuous when passing between these regions.  In what follows we also assume that there is only one region of space where the non-linear terms are important, and  that within this region there is a power-law solution to the equation of motion.  It is clear that these last two assumptions are a considerable simplification of  what may be possible,
we are clearly only studying the presence or absence of the simplest types of screening mechanism in k-essence theories.  We leave a generalisation of this to more complicated cases for future work.

There is a very clear distinction between theories where the non-linearities are purely in the potential or in the form of the coupling to matter, and those where the non-linearities rely on derivative operators.  When the non-linearities depend only on $\phi$ and not on its derivatives then the equations of motion will have the form
\begin{equation}
\Box\phi=f(\phi,\rho)\;,
\label{eq:chamsymm}
\end{equation}
where the function $f$ depends on the theory being considered.  Examples of such theories include  the chameleon and symmetron models discussed in the previous section, and in those cases $f(\phi,\rho)$ can always be locally approximated by a (potentially tachyonic) mass term.  This makes the equation of motion homogeneous, and the solutions have particularly nice properties and contain two constants of integration which are fixed by imposing boundary conditions.  In this article we focus on  terms with a non-linear dependence on $X$, and their role in the screening of a scalar force, and therefore we will not be able to make a homogeneous approximation to the equations of motion.  The power law approximation that we choose to make always allows us to find one solution to the equations of motion, but there may be others that we miss with this approach.
Systems described by equations of motions of the type given in Equation (\ref{eq:chamsymm}) have been well studied in the literature, hence in this article  we focus  on systems with kinetic non-linearities in $X$.  This means that the following analysis will not include the chameleon and symmetron cases.

Clearly many of the operators in Equation (\ref{eq:startlag}) are non-renormalisable.  In this paper we take a bottom up approach to identifying screening mechanisms, asking only; which theories are able to screen the classical scalar force?  If we do find that such screening mechanisms are possible then it will require further study to identify whether they can be embedded in a well behaved effective field theory.

\subsection{k-essence Cosmology}
We digress briefly to discuss the role of k-essence fields in cosmology.
k-essence models have been studied as explanations for the late-time acceleration of the expansion of the universe.  Commonly studied models have a separable form
\begin{equation}
\mathcal{L}\supset \sqrt{-g}K(\phi)\tilde{p}(X)\;.
\end{equation}
If the scalar field is treated as a perfect fluid, then it has equation of state
\begin{equation}
\omega_\phi =\frac{\tilde p}{2X\tilde{p}_{,X}-\tilde{p}}\;,
\end{equation}
where $\tilde{p}_{,X}$ indicates a derivative of $\tilde{p}$ with respect to $X$. Clearly with a suitably chosen $\tilde{p}$ the scalar can have an appropriate equation of state to allow it to act as dark energy.  When $\omega_\phi$ is sufficiently close to $-1$ the energy density in the scalar field will redshift away more slowly than the energy density in matter.  The instant when the scalar field comes to dominate the dynamics of the universe   is controlled by the energy scales of the higher order operators in the Lagrangian, and to solve the coincidence problem it is necessary to tune these energy scales to take values corresponding to the energy density of the universe today.
We will give a specific class of examples of this in Section \ref{sec:nonpert}.

\section{Chameleon-like screening}
\label{sec:chamlike}
Chameleon-like screening happens  when the non-linearities of the theory are only important inside the source.  To study this we consider scenarios in which only one type of non-linearity dominates.  If we  assume that the density of the source is a constant $\rho$ inside a radius $R$ and that outside there is vacuum, then for $r<R$ the equation of motion becomes
\begin{equation}
n\int_0^r dr\;r^2 \phi^{n-1}X^m +\frac{\beta M^{n+4m-4}\rho r^3}{3M_P}=mr^2 \phi^n\phi^{\prime}X^{m-1}\;.
\end{equation}

We choose a power law solution
\begin{equation}
\phi=Cr^{\alpha}\;.
\label{eq:soln}
\end{equation}
Then matching the powers of $r$ in the equation of motion requires
\begin{equation}
\alpha=\frac{2m}{2m+n-1}\;,
\end{equation}
and a full solution of  the equation of motion requires
\begin{equation}
C^{2m+n-1}=\frac{3\times2^{m-2}\beta M^{n+4m-4}M_c}{\pi\alpha^{2m}(6m+2n-3)M_PR^3}\;.
\label{eq:chamC}
\end{equation}
For the force $\phi^{\prime}$ to have a smooth behaviour at the origin we need to impose $\alpha >1$.  This requires
\begin{equation}
-|m|<n+m-1<|m|\;.
\end{equation}

As we move further away from the origin the non-linear terms become less important and we recover the linear behaviour.  We will call the radius at which we cross from the non-linear regime to the linear one $R_S$ and it must satisfy
\begin{equation}
M^{4-n-4m}\phi^nX^{m-1}|_{r=R_S}=1\;.
\label{eq:RS}
\end{equation}
Evaluated on the solution of Equation (\ref{eq:soln}), Equation (\ref{eq:RS}) determines $R_S$ in terms of the Lagrangian parameters
\begin{equation}
C R_S^{\alpha-2}= \frac{2}{\alpha^2 (6m+2n-3)}\frac{\beta \rho}{M_p}\;.
\end{equation}

Whilst still in the interior of the source object, but outside the region where non-linear terms dominate, the  solution is
\begin{equation}
\phi=\frac{\beta\rho r^2}{6M_P}+\frac{a}{r}+b\;,
\end{equation}
and  the exterior solution is
\begin{equation}
\phi=\phi_{\infty}+\frac{d}{r}\;,
\end{equation}
where $a,b,d, \phi_{\infty}$ are constants, to be determined by boundary conditions and continuity of $\phi$ and its derivative at $R$ and $R_S$.
The constant $d$ controls the strength of the force exterior to the object.  This is weaker than gravity when
\begin{equation}
\left|\frac{\beta d}{M_P}\right| \ll \frac{M_c}{8\pi M_P^2}\;.
\end{equation}
Solving the continuity equations we find
\begin{equation}
d=\frac{\beta\rho}{3M_P}(R_S^3-R^3)-C\alpha R_S^{\alpha+1}\;.
\end{equation}
Substituting in the form of $C$ from Equation (\ref{eq:chamC})
and  defining a new length scale
\begin{equation}
\tilde{R}_S^3 = R_S^3\left(1-\frac{6}{\alpha (6m+2n-3)}\right)\;,
\end{equation}
 this can be more concisely written as
\begin{equation}
d=\frac{\beta M_c}{4\pi M_P}\left(\frac{\tilde{R}_S^3}{R^3}-1\right)\;,
\end{equation}
and therefore the scalar force is weaker than gravity when
\begin{equation}
2\beta^2 \left|\frac{\tilde{R}^3_S}{R^3}-1\right| \ll 1\;.
\label{eq:thinshell}
\end{equation}
This is the k-essence version of the chameleon thin-shell condition.
It is clear that when $n,m \sim \mathcal{O}(1)$ the thin shell condition in Equation (\ref{eq:thinshell}) can only be satisfied by objects with a specific correlation between their mass and radius.  This cannot include all objects used in tests of gravity, which cover a wide range of densities and radii.  Therefore the thin shell condition in Equation (\ref{eq:thinshell}) tells us that we cannot screen the fifth force around a wide variety of objects unless $m$ and $n$ are very large.  This requires that inside the source object an operator that we would normally consider to be highly irrelevant  dominates the dynamics of the scalar field.  It is difficult to see how such a theory could be constructed in a well controlled way.

\section{Vainshtein-like Screening}
\label{sec:vainsteinoneterm}
We now  consider the Vainshtein-like situation where the non-linearities of the theory are important exterior to the source object. We can treat the massive source as a delta function $\rho(r) =M_c \delta(r)$. Assuming that only one of the non-linear terms dominates close to the matter source the equation of motion becomes
\begin{equation}
\frac{\beta M_c}{4\pi M_P}M^{n+4m -4}=m[{r}^2 \phi^n \partial_r \phi X^{m-1}]_0^r - n\int_0^r d\tilde{r} \tilde{r}^2 \phi^{n-1}X^m\;,
\label{eq:eomvainstein}
\end{equation}
Clearly the leading contribution from the right hand side of this equation must be independent of $r$ in order to balance the source on the left hand side. Therefore, away from the origin at $r=0$, the two terms on the right hand side of  Equation (\ref{eq:eomvainstein}) must have equal derivatives with opposite sign.  As we are interested in power law solutions for $\phi$, the  integral on the right hand side of Equation (\ref{eq:eomvainstein}), cannot give a constant contribution that is independent of $r$.  Therefore the leading order solution to Equation (\ref{eq:eomvainstein}) has to have  $n=0$.  If we were considering a DBI theory the condition $n=0$ would mean that the tension function $T(\phi)$ would have to remain constant over the range of $\phi$ necessary to describe the configuration around a massive source.

After setting $n=0$
we need the right-hand side of  Equation (\ref{eq:eomvainstein}) to be a constant.  Assuming that $\phi$ is well approximated by a power law
\begin{equation}
\phi=Cr^{\alpha}\;,
\end{equation}
then this occurs when
\begin{equation}
\alpha=\frac{2m-3}{2m-1}\;.
\label{alp}
\end{equation}
The equation of motion (\ref{eq:eomvainstein}) is then satisfied if
\begin{equation}
\alpha C=\left(\frac{2^{m-3}\beta M^{4(m-1)}M_c}{\pi M_P m}\right)^{1/(2m-1)}\;.
\label{C}
\end{equation}

For this to be a consistent solution to the equation of motion we need the non-linear operator we are interested in  to dominate over the canonical kinetic term in the equations motion. This requires the following consistency condition
\begin{equation}
1<mM^{4(1-m)}|X^{m-1}|\;.
\end{equation}
For the power law  solution this requires
\begin{equation}
r^{(m-1)(1-\alpha)}<R_C^{(m-1)(1-\alpha)}\equiv m^{1/2}\left(\frac{|\alpha C|}{2^{1/2}M^2}\right)^{m-1}\;.
\end{equation}

The radial force mediated by this scalar field is
\begin{equation}
F_{\phi}=\frac{\beta\phi^{\prime}}{M_P}=\frac{\alpha C\beta r^{\alpha-1}}{M_P}\;.
\end{equation}
For the moment we do not impose that the force does not diverge as $r\rightarrow 0$ because we have treated the source as a point-like object and the behaviour of the field as $r\rightarrow 0$ will be modified when the finite extent of the source is taken into account.  We will return to this in Section \ref{sec:non-pert}.
This k-essence force is weaker than gravity when
\begin{equation}
r^{\alpha+1}< R_G^{\alpha+1}\equiv \frac{M_c}{8\pi M_P \beta |\alpha C|}\;.
\end{equation}
For a successful screening mechanism  the force should be weaker than gravity close to the source  at small $r$.  Therefore we need $\alpha +1>0$. This requires
\begin{equation}
m>1 \;\;\;\;\;\mbox{ or }\;\;\;\;\; m<\frac{1}{2}\;.
\end{equation}
If we are only interested in integer values of $m$, then this makes no restriction.

So we have two length scales that control the behaviour of the field
\begin{equation}
R_G=\left(\frac{ m}{2^{m}\beta^{2m}}\right)^{1/4(m-1)}\left(\frac{M_c}{8\pi M_P M^2}\right)^{1/2}\;,
\label{rg}
\end{equation}
within which the scalar force is weaker than gravity, and
\begin{equation}
R_C=2^{1/4}m^{1/4(m-1)}\left(\frac{\beta M_c}{8\pi  M_P M^2}\right)^{1/2}\;,
\end{equation}
where the non-linear terms dominate over the canonical ones.
A successful screening mechanism  must have $R_G\leq R_C$, which requires
\begin{equation}
(2\beta^2)^{2m-1}>1\;.
\end{equation}

\subsubsection{An example model with screening}
\label{sec:example}

To illustrate more clearly how screening occurs for k-essence theories  we choose that the non-linear term that becomes dominant has $m=2$.  The theory we consider is
\begin{equation}
\mathcal{L}=-X-\frac{X^2}{M^4}-\frac{\beta\phi}{M_P}\rho\;.
\end{equation}
The resulting equation of motion is
\begin{equation}
\Box\phi+\frac{1}{M^4}\partial_{\mu}[(\partial\phi)^2\partial_{\mu}\phi]-\frac{\beta\rho}{M_P}=0\;.
\end{equation}
For a static, spherically symmetric configuration sourced by a point like object of mass $M_c$ at the origin, we can integrate the equation of motion once to find
\begin{equation}
\phi^{\prime}+\frac{1}{M^4}\phi^{\prime\;3}=\frac{\beta M_c}{4\pi M_P r^2}\;.
\end{equation}
If the non-linear term dominates close to the source then this has solution
\begin{equation}
\phi=Cr^{1/3}\;,
\end{equation}
where
\begin{equation}
C^3=\frac{3^3 \beta M^4 M_c}{4\pi  M_P}\;.
\end{equation}

The two distance scales we are interested in are
\begin{equation}
R_G=2^{-1/4}\beta^{-1}\left(\frac{M_c}{8\pi M_P M^2}\right)^{1/2}\;,
\end{equation}
\begin{equation}
R_C=2^{1/2}\beta^{1/2}\left(\frac{M_c}{8\pi M_P M^2}\right)^{1/2}\;,
\end{equation}
and the consistency condition that $R_C>R_G$ requires
\begin{equation}
2\beta^2>1\;.
\end{equation}

The ratio of the k-essence scalar force to the Newtonian force is
\begin{equation}
\frac{F_{\phi}}{F_N}=\left(\frac{r}{R_G}\right)^{4/3}=2\beta^2\left(\frac{r}{R_c}\right)^{4/3}\;.
\end{equation}
Therefore the scalar force becomes more screened as we travel further within the radius $R_G$.  This behaviour is very similar to that of the Vainshtein case considered in Section \ref{sec:vainshtein}, but k-essence screening occurs more slowly with $r$ than the Vainshtein example discussed in Section \ref{sec:vainshtein} and the screening radius has a different dependence on the couplings of the scalar field and the mass of the source.

One system within which the scalar force must be screened is the interactions between the Earth and the Moon.  We have very good experimental evidence that the orbit of the Moon around the Earth is well described by General Relativity.  Therefore if $M_c=M_{\oplus}$ then $R_G$ must be bigger than the distance from the Earth to the Moon, approximately $3\times 10^5 \mbox{ km}$.  This imposes
\begin{equation}
\beta M \lesssim \mbox { eV}\;.
\end{equation}
This is a surprisingly low energy scale.  It is intriguing that this allows for $M\sim 10^{-3} \mbox{ eV}$, the dark energy scale, suggesting that the higher order operators could be related to the currently unknown mechanism which controls the value of the cosmological constant.

\subsubsection{Can operators with non-trivial $\phi$ dependence be allowed?}
\label{sec:twoopps}

In order to satisfy equation (\ref{eq:eomvainstein}) we had to choose $n=0$.  This seems rather restrictive, and we now ask whether the scalar field is allowed to enter into the Lagrangian without derivatives if two (or more) non-linear terms become important at the same time and conspire to cancel the unwanted integral terms in the equation of motion (\ref{eq:eomvainstein}). The equation of motion is now
\begin{eqnarray}
\frac{\beta M_C}{4 \pi M_P}&=& -\int dr \;r^2(M^{4-n-4m}n\phi^{n-1}X^m+N^{4-q-4p}q\phi^{q-1}X^p)  \\
& &+[r^2\partial_r\phi(mM^{4-n-4m}\phi^nX^{m-1}+pN^{4-q-4p}\phi^qX^{p-1})]^R_0\;, \nonumber
\label{eq:twoterms}
\end{eqnarray}
where we have allowed the energy scales controlling the two nonlinear terms, $M$ and $N$ to be distinct.

We still need the integral on the right hand side of Equation (\ref{eq:twoterms}) to vanish, which means the solution must satisfy
\begin{equation}
\phi^{n-q}=-\frac{qN^{4-q-4p}}{nM^{4-n-4m}}X^{p-m}\;.
\label{eq:consistency}
\end{equation}
Then satisfying the equation of motion for a power law solution $\phi=Cr^{\alpha}$ requires us to choose
\begin{equation}
\alpha = \frac{2p-3}{q+2p-1}\;,
\end{equation}
and
\begin{equation}
C^{q+2p-1}=\frac{2^{p-3}n\beta M_c}{\pi \alpha^{2p-1}(pn-mq)M_pN^{4-q-4p}}\;.
\end{equation}

The  condition for the vanishing of the integral part of the equation of motion, given in Equation (\ref{eq:consistency}) is satisfied on these solutions if
\begin{equation}
C^{n-q-2(p-m)}r^{\alpha(n-q)-2(\alpha-1)(p-m)}=-\frac{qN^{4-q-4p}}{nM^{4-n-4m}}2^{m-p}\alpha^{2(p-m)}\;,
\end{equation}
and  the solution to this must be independent of $r$, therefore
\begin{equation}
\alpha (n-q)= 2(\alpha-1) (p-m)\;.
\label{eq:nm1}
\end{equation}
We also want the relationship between $M$ and $N$ to be independent of the mass of the source, in order to ensure that the equation of motion can be satisfied for a variety of different source objects. This means that the consistency condition has to be independent of $C$.  This requires
\begin{equation}
n-q=2(p-m)\;.
\label{eq:nm2}
\end{equation}
Unfortunately satisfying equations (\ref{eq:nm1}) and (\ref{eq:nm2}) is not possible without choosing $p=m$, which would mean that the two operators are identical.

\subsection{Multiple Operators}
\label{sec:multiple}
Let us now consider the case when the number of relevant operators is finite or even infinite, with each only depending on $X$ and not on $\phi$.
We define
\begin{equation}
M_m^{4-4m}= c_m M^{4-4m}\;,
\label{eq:mulopp}
\end{equation}
where $M$ is a characteristic scale and the Lagrangian  operators under consideration are
\begin{equation}
{\cal O}_m=c_m \frac{X^{m}}{M^{4m-4}}\;.
\end{equation}
We will analyse this situation
by singling out one term defined by its index $m$ and compare it to all the other operators. If this operator dominates then the solution to the Klein-Gordon equation is
\begin{equation}
\phi(r)= C_m r^{\alpha_m}\;,
\end{equation}
where $\alpha_m$ and $C_m$ have already been defined in Equations  (\ref{alp}) and (\ref{C}).
  We define the radii $R_{p,m}$ to describe when the $\mathcal{O}_m$ operator dominates over other terms in the equations of motion:
\begin{equation}
{\cal O}_m >{\cal O}_p \ {\rm for} \ m>p\  {\rm if} \ r<R_{m,p}\;,
\end{equation}
and
\begin{equation}
{\cal O}_m >{\cal O}_p\ {\rm for }\ p>m \ {\rm if} \ r>R_{m,p}\;.
\end{equation}
The first condition involves a finite set of operators with fewer powers of $X$ while the latter may involve an infinite number of conditions.
An expression for $R_{p,m}$ can be found using the power law solution computed in Section \ref{sec:vainsteinoneterm} when the $\mathcal{O}_m$ operator is dominant
\begin{equation}
R_{p,m}=\left(\frac{ \alpha_m C_m}{\sqrt{2} a_{pm} M^2}\right)^{(2m-1)/2}\;.
\label{eq:Rpm}
\end{equation}
For simplicity  we have defined the dimensionless combination
\begin{equation}
a_{pm}^{2(m-p)}= \frac{pc_p}{mc_m}\;.
\end{equation}
Using the expressions (\ref{alp}) and (\ref{C}) for $\alpha_m$ and $C_m$ respectively Equation (\ref{eq:Rpm}) can be rewritten as
\begin{equation}
R_{p,m}= 2^{1/4} a_{pm}^{(1-2m)/2} \left(\frac{\beta M_c}{8\pi m M_P  M^2}\right)^{1/2}\;.
\end{equation}

There is also a radius  $R_{m,G}$ given by Equation (\ref{rg}) within which the scalar force is suppressed compared to the gravitational force, this is closely related to the  Vainshtein radius discussed in Section \ref{sec:vainshtein}. The scalar force is successfully screened when
\begin{equation}
R^-_m < R_{m,G}<R_m^+\;,
\label{in}
\end{equation}
where we have defined
\begin{equation}
R^+_m= \inf_{p>m} R_{p,m}\;,
\end{equation}
and
\begin{equation}
R^-_m= \sup_{p<m} R_{p,m}\;.
\end{equation}
When this is not the case then the operator ${\cal O}_m$ does not lead to a Vainshtein mechanism. Notice too that for any given operator the Vainshtein mechanism does not extend to arbitrarily small objects: when $r<R^-_m$ the approximation breaks down and ${\cal O}_m$ is not the leading operator anymore. Moreover, nothing prevents
multiple operators ${\cal {O}}_m$ from  satisfying this inequality for objects with a size $R<R_G$. In this case, the theory which is highly non-linear has many branches. The inequality (\ref{in}) is satisfied provided
\begin{equation}
\beta_m^- <\beta < \beta^+_m\;,
\label{bet}
\end{equation}
where
\begin{equation}
\beta_m^-=2^{-1/2}\sup_{p<m}  a_{pm}^{m-1} m^{1/2(2m-1)}\;,
\end{equation}
and
\begin{equation}
\beta_m^+=2^{-1/2} \inf_{p>m}  a_{pm}^{m-1} m^{1/2(2m-1)}\;.
\end{equation}
Hence a given operator ${\cal O}_m$ leads to a Vainshtein mechanism only for a limited range of couplings. If this last inequality is not satisfied, no
Vainshtein mechanism is at play.

\subsubsection{The DBI case}

The operator expansion for the DBI model corresponds to
\begin{equation}
\mathcal{L}=\sum_{m=0}^{\infty}\left(\begin{array}{c}\;.
\frac{1}{2}\\
m
\end{array}\right)\frac{(\partial\phi)^{2m}}{[T(\phi)]^m}-V(\phi)+T(\phi)+\frac{\phi\rho}{M}
\end{equation}
As we have seen there is no Vainshtein-like screening mechanism when there is a non-trivial dependence on $\phi$.  Therefore the Vainshtein mechanism is only compatible with models where both $T(\phi)$ and $V(\phi)$ are constant over the range of field values probed by the experiment.  In this case setting $M=T^{1/4}$ we have a theory of the form given in Equation (\ref{eq:mulopp})

\begin{equation}
c_{m}=\frac{1}{m!} \frac{1}{2} \left(\frac{1}{2}-1\right)\dots \left(\frac{3}{2} -m\right)=(-1)^m\frac{\Gamma(m-\frac{1}{2})}{\Gamma(-1/2)\;. \Gamma(m+1)}
\end{equation}
It can be easily seen that $a_{pm}$  is a decreasing function of $p$ for any $m$. This implies that
\begin{equation}
R^-_m= R_{m-1,m}\;,
\end{equation}
and
\begin{equation}
R^+_m=  R_{m+1,m}\;.
\end{equation}
Now we find that
\begin{equation}
\beta_m^-=2^{-1/2}  a_{2m}^{m-1} m^{1/2(2m-1)}\;,
\end{equation}
and
\begin{equation}
\beta_m^+=2^{-1/2} \lim_{p\to \infty}  a_{pm}^{m-1} m^{1/2(2m-1)}\;.
\end{equation}
Therefore
\begin{equation}
\beta_{m}^->\beta_m^+\;,
\end{equation}
which contradicts the requirement of Equation (\ref{bet}) and so there is no Vainshtein-like screening  mechanism  at play for DBI theories.

\subsection{Non-perturbative treatment}
\label{sec:non-pert}
With the understanding we have gained from studying k-essence theories operator by operator we can now extend this to a non-perturbative study.  We will also be able to allow the source object to have finite extent.  Requiring that the solution be smooth inside as well as outside the source will be even more restrictive on the allowed set of solutions.
The previous considerations have taught us that the Vainshtein mechanism can only exist when the leading non-linear terms in the Lagrangian have no non-trivial $\phi$ dependence.  So we start from a Lagrangian
\begin{equation}
{\cal L}= M^4 f\left(\frac{2X}{M^4}\right)-\frac{\beta \phi}{M_P}T\;,
\label{eq:flag}
\end{equation}
where $f$ is an arbitrary function. In particular, this requires that any potential terms for quintessence must be subdominant over the range of field values probed by gravitational experiments.  For these theories, the equations of motion
in a spherical setting read simply
\begin{equation}
\frac{d}{dr} \left(r^2 \frac{d\phi}{dr} f'\right)= \frac{\beta \rho\Theta(r-R)}{2 M_{P}} r^2\;,
\end{equation}
where $f'=\frac{df (y)}{dy}$.  This can be integrated to give
\begin{equation}
\frac{d\phi}{dr}f^{\prime}\left(\frac{1}{M^4}\left[\frac{d\phi}{dr}\right]^2\right)=\frac{\beta M_C}{8\pi M_P}\left\{\begin{array}{cr}
\frac{r}{R^3}, & \;\;\;\;\;r<R\;, \\
\frac{1}{r^2}, & \;\;\;\;\;r>R\;,
\end{array}\right.
\label{eq:profile}
\end{equation}
where we have assumed that $(d\phi/dr)f^{\prime}$ is continuous at $r=R$ and that the solution does not diverge as $r\rightarrow 0$.

It is easier to study solutions of this equation in terms of dimensionless variables, therefore we define
\begin{equation}
x=\frac{r}{R}\;,
\end{equation}
and
\begin{equation}
\phi =M^2 R u(x)\;.
\end{equation}
We also define the distance scale
\begin{equation}
R_\star= \left(\frac{\beta M_c}{ 8\pi M_PM^2}\right)^{1/2}\;,
\end{equation}
which plays an important role in determining the dynamics of the scalar field solution.
For any continuous $f$, one can invert (\ref{eq:profile}) and find a function $g(y)$ that satisfies
\begin{equation}
g(y)f'(g^2(y))=y\;,
\end{equation}
where $g=d u /dx$ and $y=(R_\star/R)^2 x$.
Notice that $g$ is an odd function of $y$ and can always be written
\begin{equation}
g(y)= y \tilde g(y^2)\;.
\label{eq:tildeg}
\end{equation}
This will be useful later when we discuss the cosmology of these theories.
Once this inverse has been identified we can solve for the profile inside the object, $r<R$
\begin{equation}
u= u(0) +\int_0^x g\left(\left[\frac{R_\star}{R}\right]^2 \tilde x\right) d\tilde x\;.
\end{equation}
Outside the object, $r>R$, the profile is
\begin{equation}
u=  u(0) +\int_0^1 g\left(\left[\frac{R_\star}{R}\right]^2 \tilde x\right) d\tilde x+ \int_1^x  g\left(\left[\frac{R_\star}{R}\right]^2\frac{1}{\tilde x^2}\right)d\tilde x\;.
\end{equation}

Outside the object the scalar force is weaker than the gravitational force
provided
\begin{equation}
\frac{du}{dx}\le \frac{1}{\beta^2} \left(\frac{R_\star}{R}\right)^2\frac{1}{x^2}
\end{equation}
or equivalently
\begin{equation}
g\left(\left[\frac{R_\star}{R}\right]^2\frac{1}{x^2}\right)\le \frac{1}{\beta^2} \left(\frac{R_\star}{R}\right)^2\frac{1}{x^2}\;.
\label{eq:gsuppressed}
\end{equation}

\subsubsection{Non-perturbative examples}

First of all, we consider the power law case discussed earlier in Section \ref{sec:vainsteinoneterm}.  If
\begin{equation}
f(x)=  x + \frac{x^m}{2^m}\;,
\end{equation}
when the non-linear terms dominate, we find
\begin{equation}
g(y)= \left(\frac{2^m y}{m}\right)^{1/(2m-1)}\;,
\end{equation}
which gives rise to
\begin{equation}
\phi(r)=M^2 R\left(\frac{ R_\star^2}{mR^2}\right)^{1/(2m-1)} \left(\frac{r}{R}\right)^{(2m-3)/(2m-1)}\;.
\label{po}
\end{equation}
This nicely recovers  the solution we obtained previously in Equations (\ref{alp}) and (\ref{C}).  Equation (\ref{eq:gsuppressed}) applied to this choice of $f$ reproduces the expression for $R_G$,  the radius within which the scalar force is suppressed with respect to  gravity, from Equation (\ref{rg}).
From Equation (\ref{eq:profile}) it is clear that $d\phi /dr \rightarrow 0$ as $r\rightarrow 0$.  Therefore in the very center of the massive object the linear terms must dominate the behaviour of the scalar field.  However as $d\phi/dr $ grows rapidly the non-linear terms soon come to dominate the behaviour of the field, and continue to do so until $r\sim R_{\star}$.

As a second example we take DBI with
\begin{equation}
f(x)= \sqrt{1+x}\;,
\end{equation}
implying that
\begin{equation}
g(y)=\frac{2y}{\sqrt{1-4y^2}}\;.
\end{equation}
The  solution inside the source is
\begin{equation}
\phi(r)=M^2  \frac{R^3}{2R_\star^2}\left(1-\sqrt{1-4\frac{R_\star^4 r^2}{R^6}}\right)\;.
\end{equation}
When $R_\star >R$  the scalar force, which is proportional to $d \phi/dr$, will diverge at
\begin{equation}
r_{\rm divergence} = R\left(\frac{R}{\sqrt2 R_\star}\right)^2\;,
\end{equation}
which lies inside the source object.  Clearly this is not a physically acceptable situation.  Therefore we again conclude that there is no screening mechanism  for DBI models.

If we extend this analysis to models of the form
\begin{equation}
f(x)=\sqrt{1+x^m}\;,
\end{equation}
 we have now
\begin{equation}
\frac{mg^{2m-1}}{2\sqrt{1+g^{2m}}}=y\;.
\end{equation}
If there were singularities present as in the DBI example these would correspond to a divergence of $g$.  In the case when $g \gg 1$ we have
\begin{equation}
\frac{m}{2}g^{m-1}=y\;.
\label{eq:DBIesquediv}
\end{equation}
When $m\neq 1$ it is not possible for $g$ to diverge at finite $y$, which we can equivalently state as there are no divergences in the scalar force at finite $r$.  It is clear from Equation (\ref{eq:DBIesquediv}) why this does not apply to the DBI example with $m=1$.  We conclude that for models with $m>1$ it is possible to have a successful screening mechanism.

Finally we analyse a class of models related to DBI that will prove to have an interesting cosmology
\begin{equation}
f=\sqrt{1+h\left(\frac{2X}{M^4}\right)}\;,
\label{eq:models}
\end{equation}
where $h(x)= \sum_{i=1}^m{c_i}x^i$.  From Equation (\ref{eq:profile}) we find that  $d\phi/dr \rightarrow 0$  as $r\rightarrow 0$, implying that deep inside a dense body
\begin{equation}
f(x)\sim \sqrt{1+c_1 x}.
\end{equation}
As we have shown for the DBI case, this approximation breaks down inside the body as $d\phi/dr$ increases unboundedly. When this happens the higher order terms in the polynomial $h(x)$ start dominating and we can approximate
\begin{equation}
f(x) \sim \sqrt{c_m} x^{m/2}.
\end{equation}
This happens as soon as $\phi'\gg M^2$ and is guaranteed (see (\ref{po})) to be the case  as long as $R_\star \gg R$ and $r\le {\cal O}(R_\star)$. Close to $R_\star$, operators of lower order begin to  dominate again.
In this case, the screening mechanism operates for distances larger than $R$ and smaller than $R_G$ which is of the same order of magnitude as $R_\star$.
In the following section we will show that if we want this scalar field to be connected to the late-time acceleration of the expansion of the Universe then $M\sim 10^{-3}$ eV.  This implies that solar system objects are all screened in these k-essence models.

\subsubsection{Cosmology of the non-perturbative models}
\label{sec:nonpert}
We now digress to discuss the  cosmology of the k-essence models that allow for a Vainshtein-like screening mechanism.  Taking a flat FRW metric
\begin{equation}
ds^2= -dt^2 + a^2 d\vec{x}^2\;,
\end{equation}
the equation of motion for a homogeneous scalar field described by Equation (\ref{eq:flag}) can be expressed after integration as
\begin{equation}
\dot \phi f'\left(-\frac{\dot\phi^2}{M^4}\right)= -\frac{\beta \rho_m}{2M_{P}} t\;,
\label{KG}
\end{equation}
where we have assumed that $\dot \phi$ does not diverge as the scale factor goes to zero in the early Universe and,  because only non-relativistic matter contributes to the scalar equation of motion, we can assume that $\rho_m \sim a^{-3}$.
The solution to this equation can be obtained as the analytic continuation $\dot\phi/M= -i g(-i\beta \rho_m t/2M M_{P})$ or equivalently
\begin{equation}
\dot \phi= -\frac{\beta M \rho_m t}{2M_{P}}\tilde g\left(-\left[\frac{\beta \rho_m t}{2M M_{P}}\right]^2\right)\;,
\end{equation}
where $\tilde{g}$ was defined in Equation (\ref{eq:tildeg}).
The Friedmann equation for $H=\frac{\dot a}{a}$ then determines  the whole cosmological evolution
\begin{equation}
H^2= \frac{1}{3 M_{P}^2} \left(\rho_\phi + \left[1+ \frac{\beta \phi}{M_{P}}\right]\rho_m +\rho_r\right)\;,
\end{equation}
where $\rho_r$ is the energy density of all the relativistic species and the energy density of the scalar field is
\begin{equation}
\rho_\phi= 2 \dot\phi^2 f'\left(-\frac{\dot \phi^2}{M^4}\right) +M^4 f\left(-\frac{\dot \phi^2}{M^4}\right)\;.
\end{equation}
We are particularly interested in the late time evolution of the Universe. This can be conveniently analysed using the equation of state
\begin{equation}
w_\phi= \frac{p_\phi}{\rho_\phi}\;,
\end{equation}
where
\begin{equation}
p_\phi= -M^4f\left(-\frac{\dot \phi^2}{M^4}\right)\;.
\end{equation}
Using the equation of motion (\ref{KG})
we find that
\begin{equation}
w_\phi=-\frac{M^4f(-\frac{\dot \phi^2}{M^4})}{- \frac{\dot \phi \beta \rho_m t}{M_{P}} +M^4 f(-\frac{\dot \phi^2}{M^4})}\;.
\end{equation}
The scalar field dynamics mimic a cosmological constant $w_\phi \approx -1$ when
$\frac{\dot \phi \beta \rho_m t}{M_{P}} \ll M^4 f(-\frac{\dot \phi^2}{M^4})$.

Let us give a class of models leading to  late time acceleration.  We choose to start with a theory of the form
\begin{equation}
f=\sqrt{1+h\left(\frac{2X}{M^4}\right)}\;,
\end{equation}
where $h(x)= \sum_{i=1}^m{c_i}x^i$, such theories were discussed in the previous Section.  We assume that in the recent universe, $X\ll M^4$,  then the long time solution to the equation of motion is
\begin{equation}
\dot \phi= -\frac{\beta \rho_m t}{2c_1 M_{P}}\;.
\end{equation}
In this case the equation of state becomes
\begin{equation}
w_\phi= -\frac{1}{ \frac{ \beta^2 \rho_m^2 t^2}{2c_1M^4 M_{P}^2} +1}\;.
\end{equation}
We are interested in when the scalar equation of state approaches that of a cosmological constant with $w=-1$.  We define  the time $t_{\phi}$ to be when the equation of state of the scalar field is $w=-0.99$ in the recent past of the Universe, and find that it is given by the expression
\begin{equation}
\frac{t_\phi}{t_0} \sim 7.1\frac{ \beta \rho_{m0} t_0}{\sqrt{c_1} M^2  M_{P}}\;,
\end{equation}
or equivalently
\begin{equation}
t_{\phi}  \sim  9.5\frac{ \beta M_P}{\sqrt{c_1}M^2}\;.
\end{equation}
When this happens the energy density becomes $\rho_\phi \sim M^4$ which we want to  identify with $3\Omega_\Lambda H_0^2 M_{P}^2$ in order that the scalar field explains the late time acceleration of the expansion of the universe.
Therefore
\begin{equation}
\frac{t_\phi}{t_0}\sim \frac{10 \beta}{\sqrt{c_1}}\frac{\Omega_{m0}}{\sqrt{\Omega_{\Lambda}}}\;,
\end{equation}
where we have set $t_0H_0\sim 1$ and ignored other order one factors.  Imposing that this happens for a redshift of order one requires
\begin{equation}
c_1 \sim 2^6 \beta^2\;.
\end{equation}

We also need to check that the presence of this scalar field does not disrupt the cosmology of the early universe.
For the same reasons that the higher order terms in the Lagrangian dominate inside the dense sources used in fifth force experiments,  the higher order terms will also dominate in the early universe.
This gives the following equation of motion for the scalar field at early times
\begin{equation}
\sqrt{c_m}\left(\frac{\dot \phi}{M^2}\right)^{m-1}= -\frac{\beta \rho_m t }{mM^2 M_{P}}\;,
\end{equation}
where $m=2p$, $c_m>0$ and $p$ needs to be  odd to guarantee that the kinetic energy in the Lagrangian should be positive.
In what follows we will specialise to the case $m=2$ for algebraic simplicity, but the full equations are straightforward to derive.

In the early universe the ratio of the scalar density to the matter density is
\begin{equation}
\frac{\rho_{\phi}}{\rho_m}=\frac{\beta^2 \rho_m t^2}{4\sqrt{c_2} M_P^2}\;.
\end{equation}
During the matter dominated era $\rho_m \sim t^{-2}$ and so the ratio of scalar and matter densities remains constant and less than one if
\begin{equation}
\Omega_{0m}\beta^2 <\sqrt{c_2}\;.
\end{equation}
Close to the current time the higher order operators in the scalar field Lagrangian become smaller and no longer dominate the scalar field dynamics,  then the field starts to behave as a dark energy candidate as discussed above.

At even earlier times during radiation domination, the matter energy density scales as $\rho_m \sim t^{-3/2}$ therefore $\rho_m t^2 \sim t^{1/2}$ is an increasing function of time.  If the scalar field energy density is subdominant during the matter epoch then it is also forced to be subdominant in the radiation dominated epoch.

On top of having a time dependent equation of state which departs from -1 in the past of the Universe, the effects of the scalar field on the growth of structure could also be interesting. Indeed for clusters of masses between $M_c=10^{12} M_\odot$ and $M_c=10^{15} M_\odot$, the scale
\begin{equation}
R_{\star}=\left(\frac{\beta M_c}{8 \pi M_{P} M^2}\right)^{1/2}\;,
\end{equation}
which gives the order of magnitude for scales below which the fifth force is screened, varies from $0.5$ Mpc to $15$ Mpc (for $\beta/8\pi  \sim 1 $) implying that the effect of the scalar field could be felt on the largest scales. The study of this phenomenon  is left for future work.

\section{Conclusions}
\label{sec:conclusions}
In this article we have asked whether k-essence theories possess a screening mechanism.  We have been unable to find a chameleon-like mechanism
where the non-linear terms are only important in the interior of a massive source, however we find that it is possible for k-essence models to have a Vainshtein-like screening mechanism for which the non-linear terms are important exterior to the source.  However screening is not possible for all k-essence models, and in particular such a screening mechanism does not exist for the well motivated DBI model.  k-essence models that are suitable for screening are dominated by operators which have no $\phi$ dependence.  For models that do screen we find that the non-linear screening dynamics become important on distance scales below
\begin{equation}
R_{\star}=\left(\frac{\beta M_c}{8\pi M_P M^2}\right)^{1/2}\;,
\end{equation}
and that within this radius the fifth force mediated by the scalar field rapidly become smaller than the gravitational force.

In addition we find that there are models where the fifth forces can be screened, which also give rise to a cosmology that matches the one that we observe in our universe.  The same non-linearities that screen the scalar fifth force in the presence of massive sources, also cause the effects of the scalar field to be suppressed in the early universe.  At late times the equation of state can approach $w=-1$ and so the scalar field behaves like a cosmological constant term.  Recovering the observed cosmology requires tuning one of the parameters in the scalar theory to the scale of the observed cosmological constant and we have nothing to add to the discussion of how such a hierarchy of scales could be achieved.   We have shown that   a k-essence model controlled by such low energy scales can  have a successful screening mechanism  to hide from fifth force searches and a successful background cosmology. At the perturbative level, we expect effects to appear in  the Mpc range with possible observational consequences. This is left for future work.

\section*{Acknowledgments}
PB is supported in part by the Agence Nationale
de la Recherche under Grant ANR 2010 BLANC 0413 01.
CB is supported by a University of Nottingham Anne McLaren Fellowship, ACD is supported in part
by STFC.

	\bibliographystyle{JHEPmodplain}
	\bibliography{kessence_submit}

\end{document}